\documentclass[10pt,twocolumn,oneside,a4paper,final]{IEEEtran}

\IEEEoverridecommandlockouts 

\makeatletter
\let\NAT@parse\undefined
\makeatother

\usepackage{amsmath,amssymb,amsfonts,amsthm,commath,esint}
\usepackage[numbers,sort&compress]{natbib}
\usepackage{mathtools}
\usepackage[nolist,nohyperlinks]{acronym}
\usepackage{color}
\usepackage{multirow}
\usepackage{enumitem}
\usepackage{amstext}
\usepackage{cuted}
\usepackage{algorithm}
\usepackage{algorithmic}
\usepackage{microtype}
\usepackage{bm}
\usepackage{siunitx}
\usepackage{setspace}

\usepackage{indentfirst}
\usepackage{mathrsfs}
\usepackage{multicol}
\usepackage{stfloats}
\usepackage{mathtools}

\usepackage{graphicx}
\usepackage{epstopdf}

\ifCLASSOPTIONcompsoc
    \usepackage[caption=false, font=normalsize, labelfont=sf, textfont=sf]{subfig}
\else
\usepackage[caption=false, font=footnotesize]{subfig}
\fi

\makeatletter
\def\blfootnote{\xdef\@thefnmark{}\@footnotetext}
\makeatother

\begin{acronym}
\acro{BS}{base station}
\acro{CMS}{canonical minimum scattering}
\acro{DMA}{dynamic metasurface antenna}
\acrodefplural{DMA}[DMAs]{dynamic metasurface antennas}
\acro{LIS}{large intelligent surface}
\acro{LoS}{line-of-sight}
\acro{mMIMO}{massive multiple-input multiple-output}
\acro{RF}{radio-frequency}
\end{acronym}

\begin{document}

\title{Tensor-based Channel Tracking for RIS-Empowered Multi-User MIMO Wireless Systems}

\author{\IEEEauthorblockN{Jide Yuan\IEEEauthorrefmark{1},
George C. Alexandropoulos\IEEEauthorrefmark{2}, Eleftherios Kofidis\IEEEauthorrefmark{3}, \\Tobias Lindstrøm Jensen\IEEEauthorrefmark{1},
and Elisabeth De Carvalho\IEEEauthorrefmark{1}}\\
\IEEEauthorblockA{\IEEEauthorrefmark{1}Department of Electronic Systems, Aalborg University, Aalborg, Denmark}\\
\IEEEauthorblockA{\IEEEauthorrefmark{2}Department of Informatics and Telecommunications, National and Kapodistrian University of Athens, Greece}
\IEEEauthorblockA{\IEEEauthorrefmark{3}Department of Statistics and Insurance Science, University of Piraeus, Greece}
}
\maketitle

\pagestyle{empty}
\thispagestyle{empty}

\begin{abstract}
The accurate estimation of Channel State Information (CSI) is of crucial importance for the successful operation of Multiple-Input Multiple-Output (MIMO) communication systems, especially in a Multi-User (MU) time-varying environment and when employing the emerging technology of Reconfigurable Intelligent Surfaces (RISs). Their predominantly passive nature renders the estimation of the channels involved in the user-RIS-base station link a quite challenging problem. Moreover, the time-varying nature of most of the realistic wireless channels drives up the cost of real-time channel tracking significantly, especially when RISs of massive size are deployed. In this paper, we develop a channel tracking scheme for the uplink of RIS-enabled MU MIMO systems in the presence of channel fading. The starting point is a tensor representation of the received signal and we rely on its PARAllel FACtor (PARAFAC) analysis to both get the initial estimate and track the channel time variation. Simulation results for various system settings are reported, which validate the feasibility and effectiveness of the proposed channel tracking approach.
\end{abstract}

\begin{IEEEkeywords}
Channel estimation, tracking, multi-user MIMO, PARAFAC, reconfigurable intelligent surface.
\end{IEEEkeywords}

\section{Introduction}
\begin{spacing}{1}
The technology of Reconfigurable Intelligent Surfaces (RISs) is currently considered as one of the most promising technologies for future wireless communications, due to its inherent capability to offer programmable signal propagation conditions~\cite{RISE6G_COMM}.
RISs are artificial surfaces containing massive numbers of almost zero power consumption elements that have tunable impedance properties, whose joint configuration can dynamically adjust the phase of the impinging signals and thus shape the channel incurred~\cite{alexandg_RIS_scattering2021}. The tunable signal reflections can realize more directive beams towards the desired receiver (and cancel the interference to undesired ones), resulting in improved Signal-to-Interference-plus-Noise Ratio (SINR) \cite{RISE6G_COMM}, or enable spatio-temporal signal focusing in rich scattering environments~\cite{alexandg_RIS_scattering2021}.

For the RIS-based configuration to deploy its full potential, Channel State Information (CSI) for the end-to-end wireless link is required. Its estimation is, in general, a challenging task due to the absence of baseband processing capability of the (baseline) RISs~\cite{Lin2021};
most available RIS designs only contain basic circuitry for low-resolution phase control, that is, they lack the ability to also transmit training signals~\cite{alexandg_RIS_scattering2021}. Moreover, the size of the Channel Estimation (CE) problem increases with the number of the RIS elements, and of course the number of antennas and users in a Multi-User (MU) Multiple-Input Multiple-Output (MIMO) system, rendering it a very challenging task in practically large RIS-empowered networks. This challenge has already motivated intensive research and the related literature is quite extensive~\cite{RIS_CE}. An additional challenge comes from the fact that the wireless propagation environment is, in general, time-varying in practice, resulting in channel matrices that are correlated in the temporal dimension. The latter aspect has only recently started to be considered in the RIS literature, also for channel estimation \emph{and tracking} purposes. Related works address the problem via Doppler effect estimation and mitigation~\cite{zhou21,hzz21,sy21,wln21} and/or state-space modeling and associated Kalman filtering~\cite{mpl21,czlzcq21}.

Exploiting the tensor representation and a respective decomposition for the received signal constitutes a recently developed research line in the area of CE for RIS-empowered systems, inspired from the high representation and recovery potential of the tensor decomposition models~\cite{sdfhpf17}, which are highly relevant here in view of the intrinsically high-dimensional nature of the signals involved in such systems. Indeed, the received signal --- in an uplink setting --- can be viewed as having three dimensions, namely the spatial one, corresponding to the Base Station (BS) antennas, another dimension induced by the RIS phase configuration patterns, and of course the temporal one.
Methods based on the PARAllel FACtor (PARAFAC) decomposition have thus recently been proposed for estimating all involved channel matrices in RIS-empowered MU MIMO networks; see, e.g., \cite{araujo2021,wei2021,ljmy21,agah21,gaah21} and references therein.
However, the assumption of stationarity in time was made for the channels of the system in all these works.

In this paper, we consider the channel tracking problem in the uplink of a MU MIMO system empowered by a passive RIS, assuming time-varying wireless channels and relying on an adaptive PARAFAC decomposition scheme for the signal at the BS to both get the initial estimates of the User Equipment (UE)-RIS channels and track their variations in time. Moreover, the sparse nature of the incurred channels is exploited with the aid of a Generalized Approximate Message Passing (GAMP) algorithm~\cite{rangan11}. Simulation results for various system settings are reported, which validate the feasibility and effectiveness of the proposed channel tracking approach. To the best of our knowledge, this is the first tensor-based approach to the problem. We should stress at this point that this work is only a preliminary step in this direction and is to be improved and extended through related on-going work.

\textit{Notation:} Vectors and matrices are denoted by bold lower- and upper-case symbols, respectively. Calligraphic letters are used to denote higher-order tensors. $(\cdot)^{\mathrm{T}}$, $(\cdot)^{\mathrm{H}}$, and $(\cdot)^{\dag}$ stand for the transposition,  Hermitian transposition, and pseudo-inversion, respectively. $\{\mathbf{A}\}_{j,k}$ is the $(j,k)$-th element of the matrix $\mathbf{A}$ and similarly for higher-order tensors. The notation ${\text{diag}}(\mathbf{a})$ is used to denote the diagonal matrix with the elements of the vector $\mathbf{a}$ on its main diagonal. The identity matrix of order $n$ is denoted by $\mathbf{I}_n$. The symbol $\diamond$ denotes the Khatri-Rao (columnwise Kronecker) product.  $j= \sqrt{-1}$ is the imaginary unit.  The expectation operator is denoted by $\mathbb{E}\{\cdot\}$. $\|\cdot\|_{\mathrm{F}}$ stands for the Frobenius norm. $\mathbb{C}$ is the field of complex numbers.
\end{spacing}

\section{System and Channel Models}\label{sec:systemModel}
\begin{spacing}{1}
\subsection{System model}

\begin{figure}
    \centering
    \includegraphics[width =1\columnwidth]{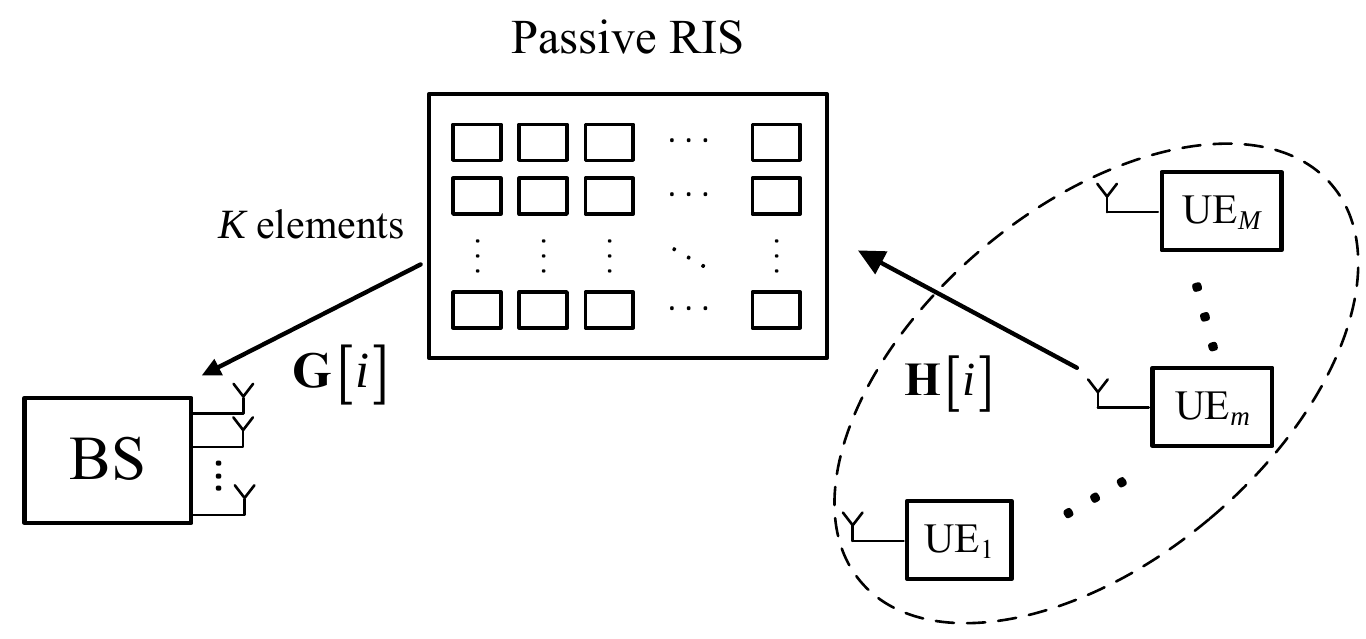}
    \caption{The considered RIS-empowered uplink multi-user MIMO system.}
    \label{fig:systemModel}
\end{figure}

We consider an RIS-empowered MU communication system in the uplink direction, as illustrated in Fig.~\ref{fig:systemModel}, where $M$ single-antenna UEs communicate with one BS equipped with $N_{\mathrm{r}}$ antennas, with the assistance of a passive RIS.
All nodes are considered synchronized and the RIS configuration control is handled by the BS~\cite{RISE6G_COMM}. The RIS contains $K$ unit elements, each with a tunable impedance capable of configuring its reflection coefficient~\cite{alexandg_RIS_scattering2021}. Moreover, channels are assumed to undergo flat fading. We neglect any direct link between the BS and any of the UEs and focus on the problem of estimating and tracking the RIS-UE channels. In cases where a direct link also exists, traditional methods can be deployed while keeping the RIS deactivated~\cite{wei2021,RIS_CE}.

We divide the time in \emph{slots} and consider the following transmission protocol:
\begin{enumerate}
    \item [i)] Each UE periodically sends its own pilot sequence, of length $S$, $L$ times per slot, with $L$ being the number of different phase profiles implementable in the RIS.
    \item [ii)] The RIS maintains a phase profile for the duration of a pilot sequence and changes to another when a whole pilot sequence has been received at the BS.
\end{enumerate}
We refer to the time needed to transmit one complete pilot sequence as the duration of a \emph{block}. Let $\mathbf{X} = \left[ {{\mathbf{x}}_1\, \cdots\, {\mathbf{x}}_M} \right]^{\mathrm{T}} \in {\mathbb{C}^{M \times S}}$ be the matrix of the pilot sequences sent from all UEs at each block, where ${\mathbf{x}}_m\in {\mathbb{C}^{S\times 1}}$ represents the pilot sequence of the $m$-th UE. The received pilot signal ${\mathbf{Y}}[ {i,l}]\in\mathbb{C}^{N_{\text{r}}\times S}$ at the $l$-th block of the $i$-th slot can be expressed as
\begin{align}\label{Received_signal_l}
{\mathbf{Y}}[ {i,l}] = \mathbf{G}[ i ]{\text{diag}}(\bm{\phi} [l]){\mathbf{H}}[ i ]{\mathbf{X}} + {\mathbf{W}}[ {i,l}],
\end{align}
where ${\mathbf{H}}[i]  = \left[ {{{\mathbf{h}}_1}[i], \ldots ,{{\mathbf{h}}_M}[ i ]} \right]\in {\mathbb{C}^{K \times M}}$ contains the channels of the RIS-UEs links and
${\mathbf{G}}[ i ] \in {\mathbb{C}^{{N_{\text{r}}} \times K}}$ is the BS-RIS channel at the $i$-th slot. $\bm{\phi} [l] \triangleq {[ {{e^{j{\theta_1}[l]}}\, \cdots\, {e^{j{\theta _K}[l]}}} ]^{\mathrm{T}}} \in {\mathbb{C}^{K \times 1}}$ is the chosen RIS phase profile at the $l$th block of the $i$th slot, with ${\theta _k}[l]\in [0, 2\pi)$, $k=1,2,\ldots,K$. ${\mathbf{W}}[ i,l] \in {\mathbb{C}^{{N_{\text{r}}} \times S}}$ represents the zero-mean spatially and temporally uncorrelated noise at the BS receiver.

\subsection{Channel model}

Considering that both the BS and the RS can be described as uniform linear arrays (normally a planar array for the RIS), we can write the following for the RIS-BS and UE-RIS channels~\cite{sr07}
\begin{align}\label{channel_G}
{\mathbf{G}}[ i ] &=\sum\nolimits_{p = 1}^P {{\alpha _{p,i}}{{\mathbf{a}}_{{N_{\text{r}}}}}\left( {{\psi _{p,i}}} \right){\mathbf{a}}_K^{\mathrm{H}}\left( {{\omega _{p,i}}} \right)},\\
\label{channel_h}{{\mathbf{h}}_m}[ i ] &= \sum\nolimits_{j = 1}^{{J_m}} {{\beta _{m,j,i}}{{\mathbf{a}}_K}\left( {{\varphi _{m,j,i}}} \right)},
\end{align}
where $P$ and $J_m$ denote the number of the propagation paths of the RIS-BS link and the link from the $m$-th UE to the RIS, respectively, and ${\alpha _{p,i}}$ and ${\beta _{m,j,i}}$ are the corresponding complex path gains.
The generic vector ${{\mathbf{a}}_{U}}\left( u \right)$
denotes the array steering vector ${{\mathbf{a}}_U}\left( u \right) = \left[ {1,{e^{ - j2\pi u}}, \ldots ,{e^{ - j2\pi \left( {U - 1} \right)u}}} \right]^{\mathrm{T}}$, where $U\in\{N_{\text{r}},K\}$ and $u\in\{{{\psi _{p,i}}},{{\omega _{p,i}}},{{\varphi _{m,j,i}}}\}$, with ${{\psi _{p,i}}},{{\omega _{p,i}}}$, and ${{\varphi _{m,j,i}}}$ being the corresponding directional cosines.
We make the assumption that ${\mathbf{G}}[ i]$ changes much more slowly than ${\mathbf{H}}[ i ]$ due to the fact that the RIS and the BS locations are fixed, rendering their wireless channel relatively stationary. We therefore assume that ${\beta _{m,j,i}}$ and ${{\varphi _{m,j,i}}}$ vary independently across slots, while ${\alpha _{p,i}}$, ${{\psi _{p,i}}}$ and ${{\omega _{p,i}}}$ may vary only every $I$ time slots.

\end{spacing}

\section{Tensor-based Channel Tracking}\label{sec:tracking}
\begin{spacing}{1}

\subsection{PARAFAC signal model}

Assuming the pilot sequences are long enough ($S\geq M$) to be orthogonal among the UEs, $\mathbf{X}$ can be assumed to have orthonormal rows, which then yields the following equivalent of~\eqref{Received_signal_l}:
\begin{align} \label{Ytilde}
    {\mathbf{\tilde{Y}}}[ {i,l}] = \mathbf{G}[i]{\text{diag}}(\bm{\phi} [l]){\mathbf{H}}[ i ] + {\mathbf{\tilde{W}}}[ {i,l}],
\end{align}
with $\mathbf{\tilde{Y}}[ {i,l}]\triangleq \mathbf{Y}[ {i,l}]\mathbf{X}^{\mathrm{H}}\in\mathbb{C}^{N_{\mathrm{r}}\times M}$ and similarly for $\mathbf{\tilde{W}}[i,l]$. It is readily verified~\cite{sdfhpf17} that this is the approximate rank-$K$ PARAFAC decomposition of the $N_{\mathrm{r}}\times M\times L$ tensor with frontal slices as in~\eqref{Ytilde}. The mode-1, 2, and~3 PARAFAC factors are $\mathbf{G}[i]$, $\mathbf{H}[i]^{\mathrm{T}}$, and $\bm{\Phi}\triangleq \left[\bm{\phi}[1],\bm{\phi}[2],\ldots,\bm{\phi}[L]\right]^{\mathrm{T}}\in\mathbb{C}^{L\times K}$, respectively.
This formulation can be made use of to estimate the channels at the first slot (as in, e.g., \cite{wei2021}) and subsequently feed those estimates to a channel tracking scheme that pursues the slot-by-slot variation of the UE-RIS channels.

However, the above requires the pilot sequence length, $S$, to be large enough to allow for the (right) inversion of $\mathbf{X}$. In a MU context, with a large number of users, $M$, this amounts to an unrealistically strict requirement for the training overhead. We therefore stick instead to~\eqref{Received_signal_l}, that we re-write in a more compact form as
\begin{align} \label{Received_signal_lZ}
    {\mathbf{Y}}[i,l] = \mathbf{G}[ i ]{\text{diag}}(\bm{\phi} [l])\mathbf{Z}[i] + {\mathbf{W}}[i,l], \;\; l=1,2,\ldots,L,
\end{align}
where
\begin{align} \label{Z[i]}
\mathbf{Z}[i]\triangleq \mathbf{H}[i]\mathbf{X}
\end{align}
is the $K\times S$ matrix of the pilot signals received at the RIS elements. Defining the $N_{\mathrm{r}}\times S\times L$ tensor $\mathcal{Y}$ with frontal slices as in~\eqref{Received_signal_lZ}, we then arrive at a PARAFAC model with factors $\mathbf{G}[i],\mathbf{Z}[i]^{\mathrm{T}}$, and $\bm{\Phi}$, that we can use to first get initial estimates of the unknown factors, $\mathbf{G}[1],\mathbf{Z}[1]$, before we track the time-varying $\mathbf{Z}[i]$, $i=2,3,\ldots,I-1$, with the aid of a recursive scheme to be developed later in this section. Then, capitalizing on the sparse nature of $\mathbf{H}[i]$, we recover it from~\eqref{Z[i]} with the aid of a GAMP approach.

\subsection{Uniqueness analysis}

Clearly, if the known matrix $\bm{\Phi}$ is of full column rank, which requires that $L\geq K$, the above PARAFAC decomposition is unique~\cite{sd15}. Indeed, one can then estimate the Khatri-Rao product of the two unknown factors, whereby these can be found via least-squares (LS) Khatri-Rao factorization (KRF)~\cite[Proposition~3.1]{sd15}.

Of course, the requirement that $L\geq K$ can be unrealistic for large-scale RISs. An alternative uniqueness condition, valid for the case that $\bm{\Phi}$ is not necessarily of full column rank, can be deduced from~\cite[Proposition~3.2]{sd15} and, in its generic form, and for a realistic setting where $L<K$ and $S<N_{\mathrm{r}},K$, states that $L+S-2\geq K$.

It should be stressed that these are only sufficient, not necessary conditions.

\subsection{The proposed channel tracking scheme}

\subsubsection{Initial channel estimation}

Estimates for $\mathbf{G}$ and $\mathbf{Z}$ at the fist slot can be computed in a way similar to~\cite{wei2021}, namely with the aid of the Alternating Least Squares (ALS) algorithm. In fact, since one of the PARAFAC factors is known, this should be called Bilinear ALS (BALS). The idea is to alternatingly solve the following two LS problems, fixing the one of the factors when computing the other: \begin{align}\label{G_estimate}
\hat{\mathbf{ G}}[1] &\triangleq \mathop {\text{argmin}  }\limits_{\mathbf{G}} \left\|\mathbf{Y}_{(1)}[1] - \mathbf{G}\left(\bm{\Phi}\diamond \mathbf{Z}^{{T}}[1] \right)^{{T}} \right\|_{{F}}^2,\\
\label{Z_estimate}{\mathbf{\hat Z}}[1]& \triangleq \mathop {\text{argmin} }\limits_{\mathbf{Z}} \left\|\mathbf{Y}_{(2)}[1] - \mathbf{Z}^{{T}}\left(\bm{\Phi}\diamond \mathbf{G}[1] \right)^{{T}} \right\|_{{F}}^2,
\end{align}
where $\mathbf{Y}_{(1)}[i] \triangleq \left[ \mathbf{Y}[i,1]\, \cdots \,\mathbf{Y}[i,L] \right]$ and
$\mathbf{Y}_{(2)}[i] \triangleq \left[ \mathbf{Y}^{{T}}[i,1]\, \cdots \,\mathbf{Y}^{{T}}[i,L]\right]^{{T}}$ are the mode-1 and mode-2 unfoldings of $\mathcal{Y}$, respectively.
This results in the iterative procedure of alternating between the following until convergence has been achieved:
\begin{align}\label{G_estimate_closef}
{\mathbf{\hat G}}[1] &= \mathbf{Y}_{(1)}[1]\left[\left(\bm{\Phi}\diamond \mathbf{\hat{Z}}^{\mathrm{T}}[1] \right)^{\mathrm{T}} \right]^\dag,\\
\label{Z_estimate_closef}{\mathbf{\hat Z}}[1] &= {\left( {{\bm{\Phi }}\diamond {\mathbf{\hat{G}}}[1]} \right)^\dag }{\mathbf{Y}}_{(2)}^{\mathrm{T}}[1].
\end{align}
Alternatively, LS~KRF may be employed.

\subsubsection{Recursive channel tracking}

Recall here that $\mathbf{G}[i+1] \approx \mathbf{G}[i]$, over a period of $I$ slots. However, $\mathbf{H}[i]$ may be changing and so will $\mathbf{Z}[i]$. Once $\mathbf{G}[i]\approx \mathbf{G}[1]$ has been estimated, an estimate for $\mathbf{Z}[i+1]$ can be simply obtained from the analog of~\eqref{Z_estimate_closef} for slot $i+1$, namely
\begin{align}\label{initial_Z}
{\mathbf{\hat Z}}[i + 1] &= \mathbf{F}^\dag\mathbf{Y}_{(2)}^{\mathrm{T}}[i+1],
\end{align}
where $\mathbf{F}$ is the $N_{\mathrm{r}}L\times K$ matrix $\bm{\Phi}\diamond\mathbf{G}[1]$. However, this is a formidable task, in view of the generally large size of this inversion problem.

It is thus preferable to follow a recursive approach to computing $\mathbf{Z}[i+1]$ based on its previous estimates. Specifically, we consider the minimization with respect to $\mathbf{Z}[i+1]$ of the following exponentially weighted squared error function at slot $i+1$:
\begin{align}\label{CostFun_F_Z}
J[i+1] \triangleq \sum_{\tau = 1}^{i + 1} \lambda ^{i + 1 - \tau} \left\|\mathbf{Y}_{(2)}^{\mathrm{T}}[ \tau  ]-\mathbf{F}[i+1]\mathbf{\hat{Z}}[\tau]\right\|_{\mathrm{F}}^{2},
\end{align}
where $\lambda\in (0,1]$ is the forgetting factor. This yields a Recursive LS (RLS) method, which resembles the classical RLS algorithm and hence its details are omitted here. It can be initialized with $\mathbf{Z}[1]$, computed in the BALS step. This two-step procedure will henceforth be referred to as BALS-RLS. 

\subsubsection{Resolution of ambiguities}

Note that the permutation ambiguity in our PARAFAC model is trivially resolved with the aid of the knowledge of $\bm{\Phi}$. Regarding the scaling ambiguities that affect the rows of $\hat{\mathbf{Z}}[i]$ and the columns of $\hat{\mathbf{G}}[1]$, they can be addressed via appropriate normalization of the channel estimates, which is an essential step for the GAMP-based channel recovery presented next.

\subsection{GAMP-based channel recovery}
\label{sec:gamp_channel_recovery}

In order to exploit the sparsity of $\mathbf{H}$ when estimating it from~\eqref{Z[i]}, we resort to a GAMP algorithm~\cite{rangan11}. This way, the channel can be recovered with fewer pilots than needed in sparsity-unaware methods. This algorithm belongs to a family of methods that are based on Gaussian and quadratic approximations of loopy belief propagation, a unified methodology of estimating sparse/dense matrices. The problem can be stated as follows:
\begin{align}\label{Recovery_H}
{\mathbf{\hat H}}[ i ] = \mathop {\text{argmin} }\limits_{\mathbf{H}} \left\| {{\hat{\mathbf{Z}}}[ i ] - {\mathbf{H}}[i]{\mathbf{X}}} \right\|_{\mathrm{F}}^2.
\end{align}
To fit the GAMP formalism, this optimization problem can be equivalently transformed into~\cite{8370683}:
\begin{align}\label{Recovery_H_new}
\hat {\mathbf{H}}_\text{a}[ i ] = \mathop {\text{argmin} }\limits_{\mathbf{H}_\text{a}} \left\| {{\hat{\mathbf{Z}}}^T[ i ] \mathbf{U}- {\mathbf{X}}^T} {\mathbf{H}_\text{a}}[i]\right\|_{\mathrm{F}}^2,
\end{align}
where ${\mathbf{H}_\text{a}}\left[ i \right]\triangleq{\mathbf{H}^T}\left[ i \right] \mathbf{U}$ is a sparse matrix representing the channel in the angular domain, with $\mathbf{U}$ being the $K\times K$ normalized Discrete Fourier Transform (DFT) matrix. This manipulation explores the row-sparsity of the channel, making it more efficient to apply GAMP for computing $\hat {\mathbf{H}}_\text{a}[ i ]$. The sought for estimate $\hat{\mathbf{H}}[ i ]$ can then be computed as $\hat{\mathbf{H}}[ i ] =\mathbf{U} {\hat{\mathbf{H}}_a^{\mathrm{T}}}[ i ]$.

GAMP performs better than other sparsity-aware methods in this task. The reason is that ${\mathbf{H}}_a[ i ]$ is not a strictly sparse matrix due to the angular spread and the resolution limit of the DFT matrix. GAMP, however, can cope with both sparse and dense matrices, and this gives it an advantage over other algorithms~\cite{Donoho09}. Its worst-case complexity per iteration is $\mathcal{O}(MS)$, yielding an overall complexity $\mathcal{O}(TMS)$, with $T$ being the (maximum) number of iterations.

\end{spacing}
\section{Simulation Results}\label{sec:simulations}

\begin{spacing}{1}
We adopt the Normalized Mean Squared Error (NMSE) at each slot $i$,
\[\text{NMSE}_{\mathbf{A}[i]} =  \mathbb{E}\left\{\frac{\left\| \mathbf{\hat{A}}[ i ] - \mathbf{A}[ i ]\right\|_{\mathrm{F}}^2}{\left\|\mathbf{ A}[ i ]\right\|_{\mathrm{F}}^{2}}\right\},
\]
with ${\mathbf{ A}}[i]\in\{{\mathbf{G}[i]\mathbf{Z}[i]},{\mathbf{H}[i]}\}$, as the performance metric in our comparative study, where the expectation is computed with averaging over 100 Monte Carlo runs.

All angles are chosen uniformly at random. The ${{\varphi _{m,j,i}}}$s are left to vary from one slot to the other, while the ${{\psi _{p,i}}}$s and ${{\omega _{p,i}}}$s are kept constant over $I$ slots.
We have adopted truncated DFT matrices for both the pilot matrix $\mathbf{X}$ and the RIS phase profile matrix $\bm{\Phi}$. There are $M=20$ users, with $J_m=4$ for all of them, while the BS is equipped with $N_{\text{r}}=16$ antennas. The forgetting factor was set to $\lambda = 0.5$.

\begin{figure}
    \centering
    \includegraphics[width = 0.85\columnwidth]{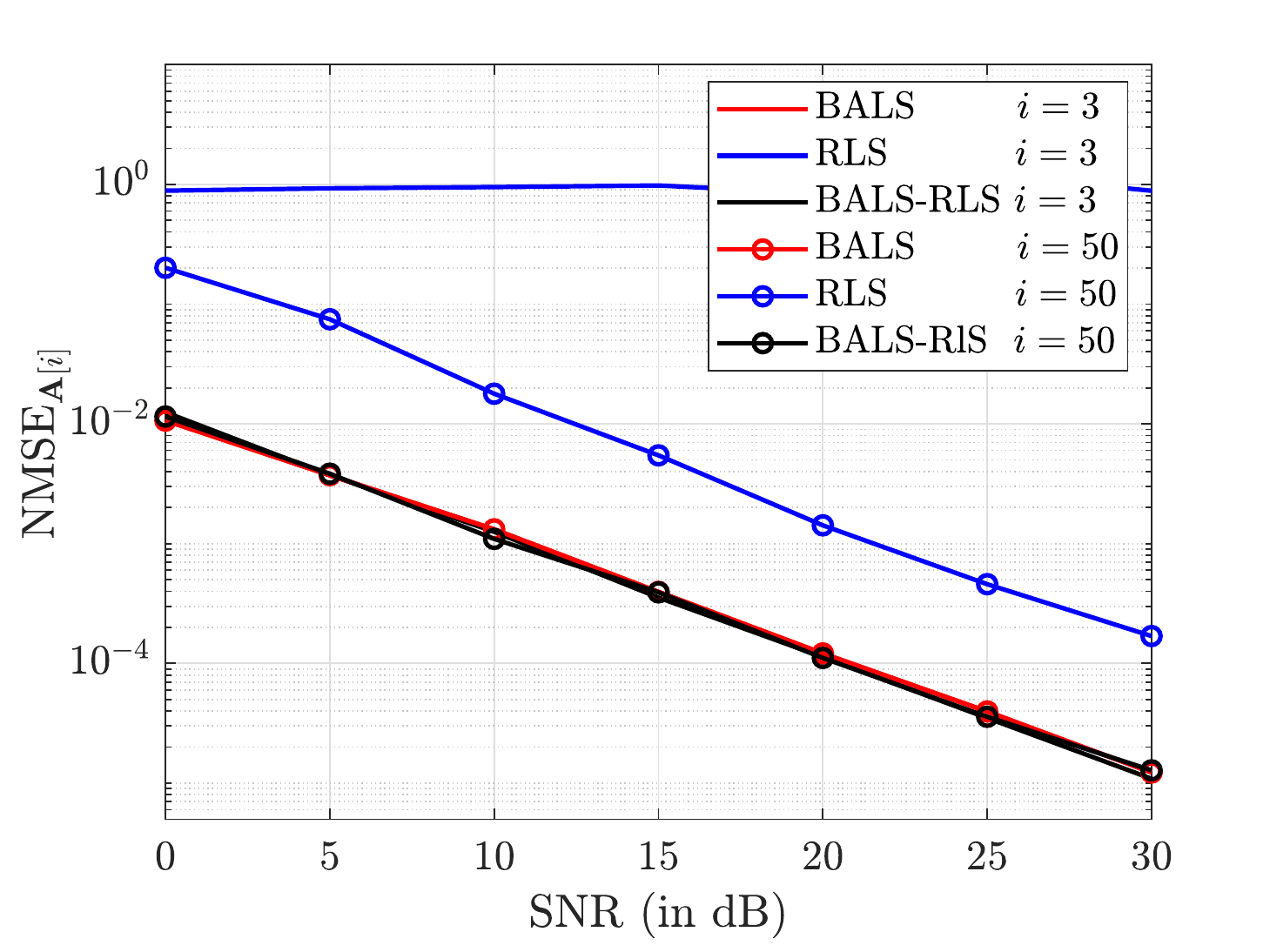}
    \caption{The NMSE performance of the BALS, RLS, and BALS-RLS algorithms versus SNR, at the $3^{\text{rd}}$ and 50-th slots. $K=64$ and $I = 100$.}
    \label{fig:SNR}
\end{figure}

\begin{figure}[!t]
    \centering
    \includegraphics[width = 0.85\columnwidth]{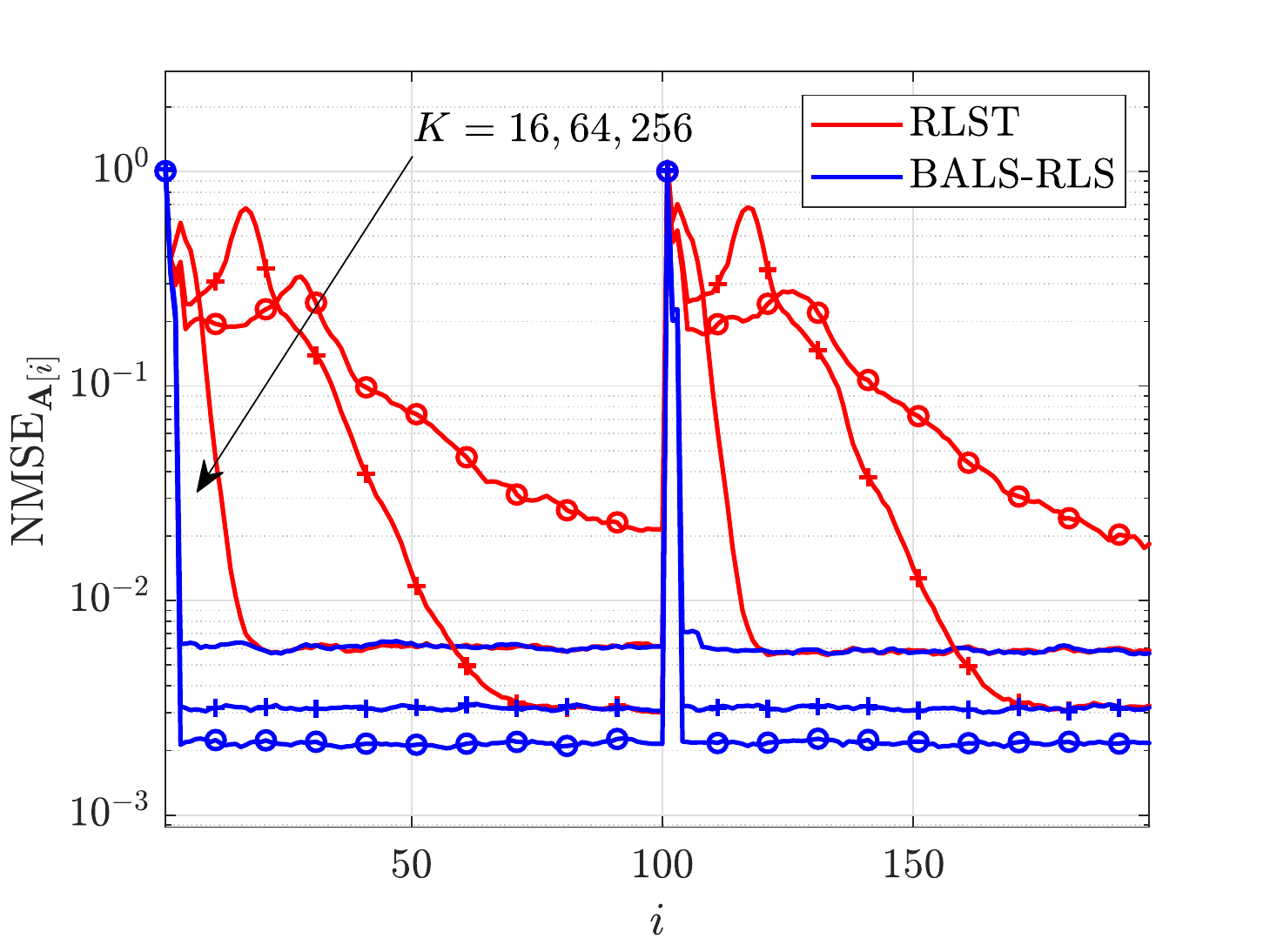}
    \caption{The NMSE performance of the RLS and BALS-RLS algorithms for $I = 100$ slots. SNR=10~dB.}
    \label{fig:Conver_iter}
\end{figure}

To evaluate the tracking ability of the BALS-RLS algorithm, we first check the NMSE for ${\mathbf{A}}[i]={\mathbf{G}[1]\mathbf{Z}[i]}$. Fig.~\ref{fig:SNR} depicts the NMSE performance as a function of the Signal-to-Noise Ratio (SNR) for three algorithms: BALS-RLS, recursive tracking with random initialization (RLS), and BALS for all time slots.
The NMSE at the 3rd and the 50-th time slots is plotted, assuming a relatively stable RIS-BS channel, with $I=100$. We can see that the BALS and BALS-RLS algorithms provide nearly the same performance in both time slots. On the contrary, the RLS algorithm performs poorly at the 3rd slot and is still not comparable to the other two algorithms at $i=50$. The NMSE performance of the two recursive schemes, RLS and BALS-RLS, is plotted in Fig.~\ref{fig:Conver_iter} as a function of the slot index, for $I=200$ time slots in total, and for various RIS sizes.
In this experiment, the RIS-BS channel changes every 100 slots. We observe that RLS and BALS-RLS achieve the same level of accuracy when $K=16$ and~64. However, RLST requires observations from more time slots to converge with larger $K$, and it does not manage to converge when $K=256$, which seriously limits its application. In contrast, the BALS-RLS algorithm converges within a few time slots in all cases.

\begin{figure}
    \centering
    \includegraphics[width = 0.88\columnwidth]{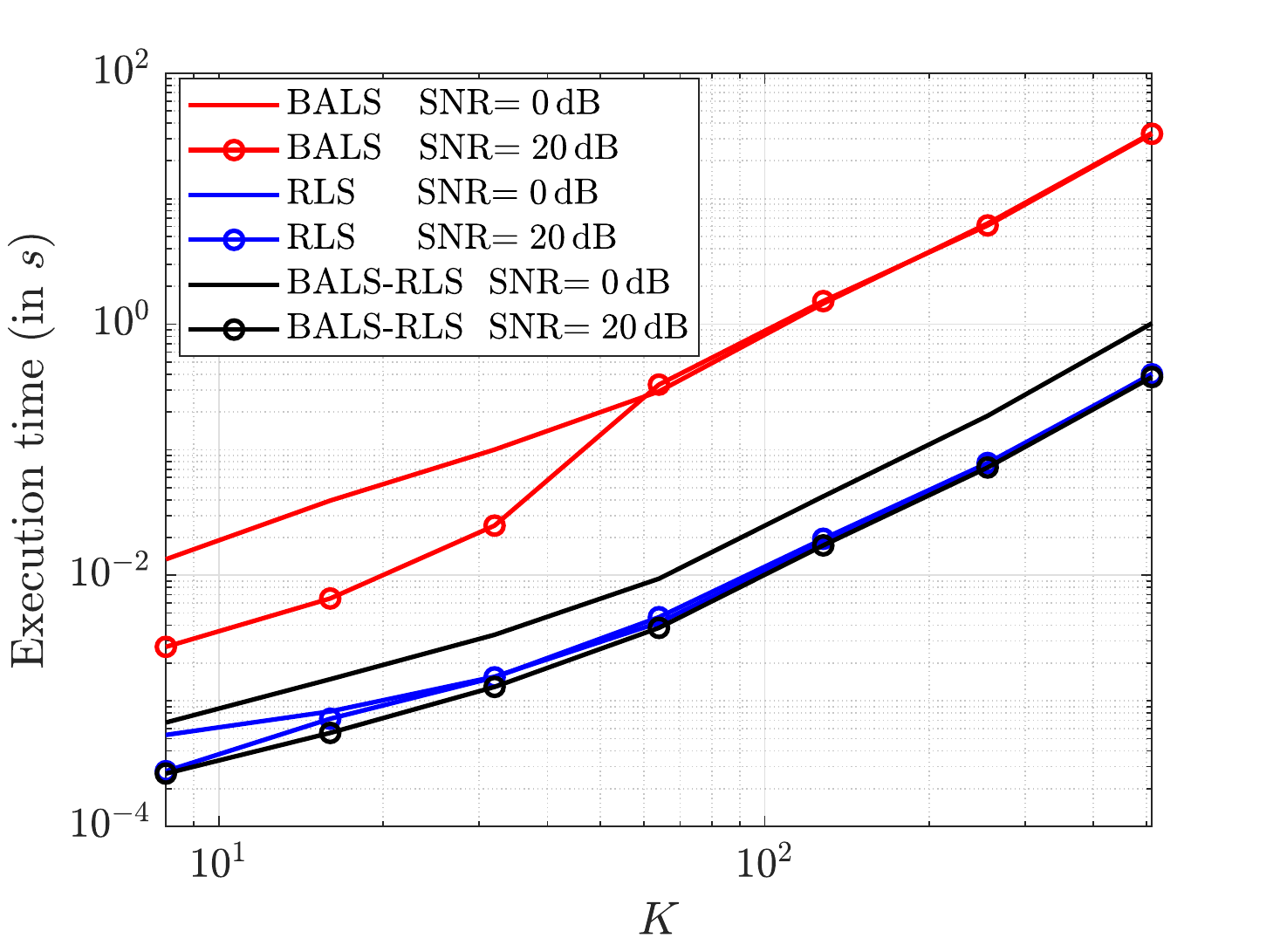}
    \caption{Comparison of the run times of the BALS, RLS, and BALS-RLS algorithms versus the number of the RIS elements, $K$.}
    \label{fig:Time}
\end{figure}

Fig.~\ref{fig:Time} plots the computational cost of the three algorithms with respect to $K$.
The curves represent the average execution time per time slot. We first observe that the cost of BALS is nearly two orders of magnitude higher than those of the other algorithms. 
The complexities of the two recursive schemes are significantly lower and very similar, especially at high SNRs.

\begin{figure}
    \centering
    \includegraphics[width = 0.88\columnwidth]{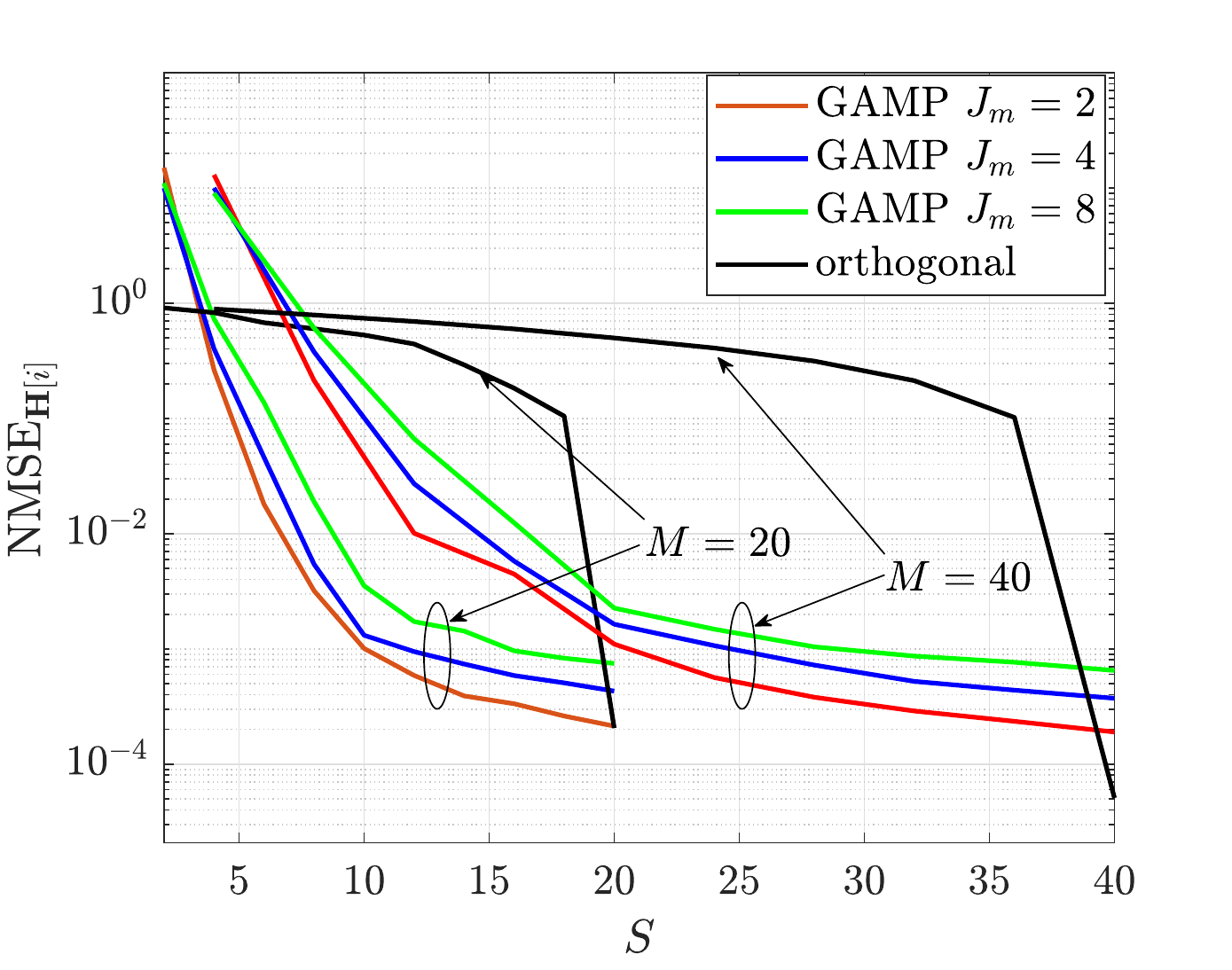}
    \caption{Comparison of GAMP and the orthogonal pilot scheme for different pilot sequence lengths. $K =64$ and SNR=0~dB.}
    \label{fig:pilot_length}
\end{figure}

We finally illustrate the recovery performance of the GAMP-based algorithm. In Fig.~\ref{fig:pilot_length}, we depict the NMSE performance of $\mathbf{H}[i]$ as a function of the pilot sequence length, $S$, for different numbers of propagation paths per UE. The orthogonal pilot scheme is included as a benchmark.
It can be observed that a relatively stable estimation accuracy is obtained by GAMP, when the number of pilots is approximately equal to $\frac{M}{2}$. Due to the angular spread effect, the channel is not a strictly sparse matrix in the angular domain. This imposes a limit on the recovery ability of GAMP and hence it performs somewhat worse than the orthogonal pilot scheme when $S=M$.

\end{spacing}

\section{Conclusion}\label{sec:conclusions}

Capitalizing on the PARAFAC decomposition of the received signal in the uplink of a RIS-empowered MU MIMO system, we developed a low-complexity algorithm for tracking the time-varying user-RIS channels. Their sparse nature has also been exploited, via a GAMP scheme. The reported simulation results showcased that the proposed tracking method can achieve the same level of estimation accuracy with its batch (BALS) version, though with a significantly lower computational and training overhead.

This preprint reports preliminary results on tensor-based RIS-empowered channel tracking, \emph{for the first time}. Extended and improved results will be reported elsewhere.



\bibliographystyle{setup/IEEEtran}

\end{document}